\documentclass[aps,prl,showpacs,twocolumn]{revtex4}
\usepackage{graphicx}

\begin{document}

\title{Anomalous reflection and excitation of surface waves in metamaterials}
\author{G. Brodin, M. Marklund, L. Stenflo and P. K. Shukla}
\affiliation{ Department of Physics, Ume{\aa } University, 
SE-901 87 Ume\aa,~ Sweden}

\date{Received 20 February 2007; accepted 1 March 2007}

\begin{abstract}
We consider reflection of electromagnetic waves from layered structures with
various dielectric and magnetic properties, including metamaterials.
Assuming periodic variations in the permittivity, we find that the
reflection is in general anomalous. In particular, we note that the specular
reflection vanishes and that the incident energy is totally reflected in the
backward direction, when the conditions for resonant excitation of leaking
surface waves are fulfilled.
\end{abstract}

\maketitle

\paragraph{Introduction}

Materials with new kinds of dielectric and magnetic properties, e.g.
left-handed metamaterials, have stimulated much recent work, see e.g. \cite
{Agranovich}. \ In the light of the anomalous electromagnetic
characteristics \cite
{Agranovich,Shalev2007,Makarov2006,Chowdhury2007,Marklund2005,Marklund2006,Ruppin2000,Shadrivov2004,Shadrivov2003,Shadrivov2005},
it is necessary to reconsider many well known problems, accounting also for
metamaterials. Due to their extraordinary properties, left-handed materials
are believed to become highly important in various technological
applications \cite{Agranovich,Shalev2007}. The first examples of
metamaterials had their left-handed properties in the microwave range, and
then later for infrared waves \cite{Nature2007}. Recently materials showing
anomalous refractive behaviour have been constructed also in the optical
regime \cite{Shalev2007}. %[add Nature-Ref].

It has previously been shown in Ref. \cite{Aliev1993} that certain layered
dielectric structures give rise to peculiar reflection phenomena, provided
surface waves \cite{Ruppin2000,Shadrivov2004,GradovStenflo1982,GradovStenflo1983,Stenflo1996}
can be excited. Such waves have previously been
considered in connection with layered structures \cite
{Shadrivov2003,Shadrivov2005,Aliev1993,Zhelyazkov1987}, open plasma wave guides \cite
{Grozev1991,Aliev2000}, and fusion plasmas \cite{Yu1986}.

In the present paper, we consider the reflective properties of a dielectric
and magnetic medium with arbitrary permittivity and permeability, which is
covered by a dielectric layer. We show that for a homogeneous medium, we
have the familiar specular reflection independently of the material
parameters. However, with periodic variations in the dielectric
permittivity, as can be found in nature \cite
{Vukusic2006,Stenflo-Yakimenko1978} and constructed in laboratories \cite
{Agranovich,Shalev2007,Nature2007}, we find that the incident energy can be
completely converted into a backscattered energy flux, a phenomenon which is
associated with the resonant excitation of leaking surface waves. The
necessary conditions for obtaining backward reflection will be deduced, and
possible applications of this peculiar mirror effect are consequently pointed out.

\paragraph{Homogeneous media}

We first consider a semi-infinite ($z>z_{0}$, region $a$) material with
arbitrary relative permittivity $\varepsilon _{a}$ and permeability $\mu
_{a} $. Specifically we will allow these constants to have arbitrary signs,
such that for example metamaterials with negative signs of both $\varepsilon 
$ and $\mu $ are included as special cases. This material is assumed to be
covered by a dielectric layer with arbitrary permittivity $\varepsilon _{b}$
and $\mu _{b}=1$ in the region $0<z<z_{0}$ (region $b$), whereas we have
vacuum at $z<0$. Furthermore, we let a $p$-polarized wave with a magnetic
field $(B_{0}\mathbf{\hat{y}}/2)\mathrm{exp}(ik_{x}x+ik_{z}z-i\omega t)+%
\mathrm{c.c}$., be incident on this structure. Here $\mathrm{c.c.}$ denotes
complex conjugate, $B_{0}$ is the wave amplitude, and $k_{x,z}>0$. We denote
the angle of incidence by $\theta $ (given by $k_{x}=(\omega /c)\sin \theta $
or $k_{z}=(\omega /c)\cos \theta $). As a prerequisite for the more
complicated calculations in the next section, we here study the fields
induced in the materials for the special case where the waves are evanescent
in both regions $a$ and $b$. The magnetic field in each of the three regions 
$z<0$, $0<z<z_{0}$ and $z>z_{0}$ is thus written in the form $\mathbf{B}=(\mathbf{%
\hat{y}}/2)B(x,z)\mathrm{exp}(-i\omega t)+\mathrm{c.c}.$ In the vacuum
region the $y$-component satisfies the equation %1
\begin{equation}
\partial _{x}^{2}B+\partial _{z}^{2}B+\frac{\omega ^{2}}{c^{2}}B=0,
\label{eq:vacuum-wave}
\end{equation}
which has solutions %2
\begin{equation}
B=B_{0}e^{ik_{x}x+ik_{z}z}+B_{0}\left(
R_{s}e^{ik_{x}x}+R_{b}e^{-ik_{x}x}\right) e^{-ik_{z}z},
\label{eq:vacuum-sol}
\end{equation}
where $R_{s}$ and $R_{b}$ are constants which represent the specularly and
backward reflected waves, respectively. Normally, one would expect $R_{b}$
to be zero. However, we note that metamaterials may behave differently
compared to normal materials. Furthermore, inclusion of the $R_b$-term will be
useful for the calculations made in the next section. From energy
conservation, evanescent waves in regions $a$ and $b$ imply $R_{b}+R_{s}=1$.
Next, in region $b$ we solve the equation %3
\begin{equation}
\partial _{x}^{2}B+\partial _{z}^{2}B+\frac{\omega ^{2}\varepsilon _{b}}{%
c^{2}}B=0,  \label{eq:wave-region-a}
\end{equation}
and write the solution in the form %4
\begin{eqnarray}
  B=\frac{B_{0}}{2}\left( 1-\frac{ik_{z}\varepsilon _{b}}{\kappa _{b}}\right)
  e^{ik_{x}x}\left( e^{-\kappa _{b}z}+re^{\kappa _{b}z}\right) 
  \nonumber \\ 
  +\left(
  R_{s}e^{ik_{x}x}+R_{b}e^{-ik_{x}x}\right) \left( e^{\kappa
  _{b}z}+re^{-\kappa _{b}z}\right) ,  
\label{eq:region-a-sol}
\end{eqnarray}
where $\kappa _{b}=(\omega /c)(\sin ^{2}\theta -\varepsilon _{b})^{1/2}$,
and where $r=(1+ik_{z}\varepsilon _{b}/\kappa _{b})/(1-ik_{z}\varepsilon
_{b}/\kappa _{b})$ represents the reflectivity at $z=0$. We note that the
standard boundary conditions expressed in terms of $B$, namely the
continuity of $B$ and $(1/\varepsilon )\partial _{z}B$ are satisfied at $z=0$
for arbitrary values of $R_{s}$ and $R_{b}$. In region $a$ we use the
equation %5
\begin{equation}
\partial _{x}^{2}B+\partial _{z}^{2}B+\frac{\omega ^{2}\varepsilon _{a}\mu
_{a}}{c^{2}}B=0,  \label{eq:wave-region-b}
\end{equation}
and write the solution as %6
\begin{equation}
B=(B_{+}e^{ik_{x}x}+B_{-}e^{-ik_{x}x})e^{-\kappa _{a}(z-z_{0})},
\label{eq:region-b-sol}
\end{equation}
with $\kappa _{a}=(\omega /c)(\sin ^{2}\theta -\varepsilon _{a}\mu
_{a})^{1/2}$. Since both $\mu $ and $\varepsilon $ vary across the
boundary, the boundary conditions are now the continuity of $B/\mu $ and $%
(1/\varepsilon )\partial _{z}(B/\mu )$. Matching the solutions in regions $a$
and $b$, we obtain %7-8
\begin{eqnarray}
  && 
  \frac{B_{0}}{2}\left( 1-\frac{ik_{z}\varepsilon _{b}}{\kappa _{b}}\right) %
  \big[ \left( e^{-\kappa _{b}z_0}+re^{\kappa _{b}z_0}\right) 
  \nonumber \\ && \qquad\qquad
  +\left( e^{\kappa
  _{b}z_0}+re^{-\kappa _{b}z_0}\right) R_{s}\big] =\frac{B_{+}}{\mu _{a}},
\label{eq:boundary-1a} \\
  &&
  \frac{B_{0}}{2}\left( 1-\frac{ik_{z}\varepsilon _{b}}{\kappa _{b}}\right)
  \left( e^{\kappa _{b}z_0}+re^{-\kappa _{b}z_0}\right) R_{b} = \frac{B_{-}}{\mu
  _{a}},  \label{eq:boundary-1b}
\end{eqnarray}
as well as %9-10
\begin{eqnarray}
  &&\!\!\!\!\!\!\!\!\!\!\!\!
  \frac{\kappa _{a}B_{0}}{2}\left( 1-\frac{ik_{z}\varepsilon _{b}}{\kappa _{b}}%
  \right) \big[ \left( -e^{-\kappa _{b}z_0}+re^{\kappa _{b}z_0}\right) 
  \nonumber \\ && \qquad\qquad
  +\left(
  e^{\kappa _{b}z_0}-re^{-\kappa _{b}z_0}\right) R_{s}\big] =-\frac{\kappa
  _{a}\varepsilon _{b}B_{+}}{\varepsilon _{a}\mu _{a}},  
\label{eq:boundary-2a} \\
  &&\!\!\!\!\!\!\!\!\!\!\!\!
  \frac{\kappa _{b}B_{0}}{2}\left( 1-\frac{ik_{z}\varepsilon _{b}}{\kappa _{b}}%
  \right) \left( e^{\kappa _{b}z_0}-re^{-\kappa _{b}z_0}\right) R_{b} = -\frac{%
  \kappa _{a}\varepsilon _{b}B_{-}}{\varepsilon _{a}\mu _{a}}.
\label{eq:boundary-2b}
\end{eqnarray}
Here we note that the equations involving $B_{-}$ couple only to the
backward reflection coefficient $R_{b}$, and that the incident wave does not
act as a source for this part. Thus, we can interpret this contribution as
a leaking surface wave \cite{Kats2007,Fourkal2007} that is localized around the boundary $z=z_{0}$ and
decays exponentially away from it, but couples to a propagating
(''leaking'') part in the boundary region. However, no matter what kind of
material properties we have in region $a$, this mode cannot be excited by
the incident wave, as the boundary conditions do not allow it to couple to
the incident wave. The field profile of the surface mode is shown in Fig. 1.
Furthermore, as will be demonstrated in the next section, the properties of
the surface wave are crucial for the reflection properties when the media in
the different regions are not homogeneous. 

%%%%%%%%%%%%%%%%%%%%%%%%Figure%%%%%%%%%%%%%%%%%%%%%%%%%%%%%%%
\begin{figure}[tbp]
\centering
\includegraphics[width=0.9\columnwidth]{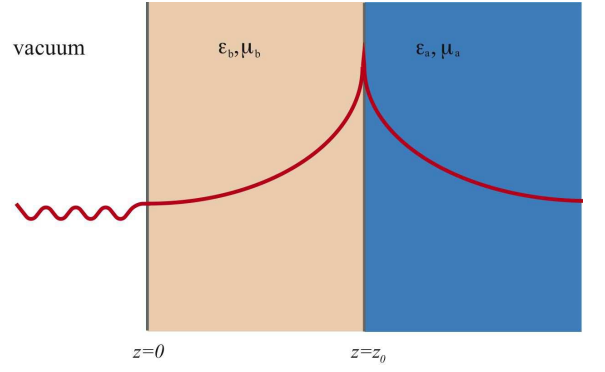}
\caption{A schematic representation of the $z$-dependence of the leaking
surface wave in the various regions.}
\end{figure}
%%%%%%%%%%%%%%%%%%%%%%%%%%%%%%%%%%%%%%%%%%%%%%%%%%%%%%%%%%%%%%

Due to the decoupling of the surface waves in a homogeneous medium, however,
we here obtain $R_{s}=1$ from the boundary conditions, and hence we always
have a purely specular reflection with this geometry. Let us stress that for
a metamaterial, the surface wave localized around $z=z_{0}$ carries energy
in the opposite direction along the x-axis, as compared to the incident
wave. However, as shown above, this does not affect the direction of the
reflected energy in the vacuum region.

\paragraph{Periodic media}

We next extend our previous analysis to consider a semi-infinite ($z>z_{0}$)
medium with a periodically varying permittivity $\varepsilon
_{a}=\varepsilon _{a0}+\varepsilon _{a1}\cos (2k_{x}x)$, where $\varepsilon
_{a0}$ and $\varepsilon _{a1}$ are constants. We note that such a medium can
be constructed in laboratories \cite{Agranovich,Shalev2007,Nature2007}, but
also that such structures are common in nature \cite{Vukusic2006}. The
medium in region $b$ is the same as in the previous section. For simplicity,
the permeability $\mu _{a}$ in region $a$ is kept constant. Generally, the
interaction between the incident wave and the periodic dielectric structure
in medium $a$ will result in the generation of new waves corresponding to
the sum and difference of the wave vectors. Here, we choose the $x$%
-component of the incident wave vector such that it is equal to half the
wave number of the periodic modulation in region $a$. This choice of
incident angle $\theta $ means that the backscatter generation mechanism is
resonant and thus much more important than the corresponding interaction
mechanism for other values of $\theta $. Since the medium and thereby the
wave equations are the same as in the previous analysis in the vacuum region
and region $b$, we can adopt the corresponding solutions (\ref
{eq:region-a-sol}) and (\ref{eq:region-b-sol}). Furthermore, for region $a$
we can closely follow the analysis in Ref. \cite{Aliev1993}. Including the
spatial dependence of the permittivity, the wave equation (\ref
{eq:wave-region-b}) is then generalized to %11
\begin{equation}
\frac{1}{\varepsilon _{a}\mu _{a}}\partial _{z}^{2}B+\partial _{x}\left( 
\frac{1}{\varepsilon _{a}\mu _{a}}\partial _{x}B\right) +\frac{\omega ^{2}}{%
c^{2}}B=0.  \label{eq:varying-permitivity}
\end{equation}
Similarly, the ansatz (\ref{eq:region-b-sol}) is generalized to %12
\begin{equation}
B=B_{+}(z)e^{ik_{x}x}+B_{-}(z)e^{-ik_{x}x}.  \label{eq:ansatz-periodic}
\end{equation}
Assuming that $\varepsilon _{a1}$ is much smaller than $\varepsilon _{a0}$,
we then obtain the two coupled equations %13
\begin{eqnarray}
  &&
  \partial _{z}^{2}B_{\pm }-\left( \kappa _{a}^{2}-\frac{\varepsilon
  _{a1}^{2}\omega ^{2}}{c^{2}\varepsilon _{a0}\mu _{a}}\right) B_{\pm }
  \nonumber \\ &&\quad
  =-\frac{%
  \varepsilon _{a1}}{2\varepsilon _{a0}\mu _{a}}\left[ \partial _{z}^{2}B_{\mp
  }-\left( k_{x}^{2}+2\kappa _{a}^{2}\right) B_{\mp }\right] ,
\label{eq:coupled}
\end{eqnarray}
with the boundary conditions %14
\begin{equation}
(B_{+})_{z=z_{0}}=\tilde{B_{0}}\left[ r+R_{s}+(1+rR_{s})e^{-2\kappa
_{b}z_{0}}\right] ,  \label{eq:BC1}
\end{equation}
%15
\begin{equation}
(B_{-})_{z=z_{0}}=\tilde{B_{0}}R_{b}\left( 1+re^{-2\kappa _{b}z_{0}}\right) ,
\label{eq:BC2}
\end{equation}
%16
\begin{eqnarray}
  &&
  (\partial _{z}B_{+})_{z=z_{0}}=\tilde{B_{0}}\frac{\kappa _{b}}{\varepsilon
  _{b}}\Big\{ \varepsilon _{a0}\mu _{a}\left[ r+R_{s}-(1+rR_{s})e^{-2\kappa
  _{b}z_{0}}\right] 
  \nonumber \\ && \qquad
  +\frac{\varepsilon _{a1}}{2}R_{b}\left( 1-re^{-2\kappa
  _{b}z_{0}}\right) \Big\} ,  \label{eq:BC3}
\end{eqnarray}
and %17
\begin{eqnarray}
  &&
  (\partial _{z}B_{-})_{z=z_{0}}=\tilde{B_{0}}\frac{\kappa _{b}}{\varepsilon
  _{b}}\Big\{ \varepsilon _{a0}\mu _{a}R_{b}\left( 1-re^{-2\kappa
  _{b}z_{0}}\right) 
  \nonumber \\ && \qquad
  +\frac{\varepsilon _{a1}}{2}\left[
  r+R_{s}-(1+rR_{s})e^{-2\kappa _{b}z_{0}})\right] \Big\} ,  \label{eq:BC4}
\end{eqnarray}
where $\kappa _{a}=(\omega /c)(\sin ^{2}\theta -\varepsilon _{a0}\mu
_{a})^{1/2}$ and $\tilde{B_{0}}=(B_{0}/2)(1-ik_{z}\varepsilon _{b}/\kappa
_{b})\mathrm{exp}(\kappa _{b}z_{0})$ The solution of (\ref{eq:coupled}) can
be found by expanding $B_{\pm }$ in powers of the small parameter $%
\varepsilon _{a1}$. After lengthy but straightforward calculations, we
obtain the solution %18
\begin{equation}
B_{\pm }\approx C_{\pm }e^{-\kappa _{a}(z-z_{0})}-\frac{\varepsilon
_{a1}(k_{x}^{2}+\kappa _{a}^{2})z}{4\kappa _{a}\varepsilon _{a0}\mu _{a}}%
C_{\mp }e^{-\kappa _{a}(z-z_{0})},  \label{eq_B-solution}
\end{equation}
where $C_{\pm }$, as well as $R_{b}$ and $R_{s}$, are complex constants that
can be found from (\ref{eq:BC1})-(\ref{eq:BC4}). As $\varepsilon
_{a1}/\varepsilon _{a0}\ll 1$, only weakly damped leaking surface waves are
of interest here. Thus, from now on we focus our attention on the regime $%
\exp (-\kappa _{a}z_{0})\ll 1$. We then obtain the simple approximate
solution %19
\begin{equation}
R_{s}\approx -r\frac{|D|^{2}}{D^{2}}\left[ 1+\frac{\varepsilon
_{a1}^{2}\omega ^{4}(1-r^{2})e^{-2\kappa _{b}z_{0}}}{8\kappa
_{a}^{4}c^{4}r|D|^{2}D}\right] ,  \label{eq:Rs-solution}
\end{equation}
and %20
\begin{equation}
R_{b}\approx \varepsilon _{a1}\frac{\omega ^{2}}{2\kappa _{a}^{2}c^{2}}\frac{%
(1-r^{2})}{D^{2}}e^{-2\kappa _{b}z_{0}},  \label{eq:Rb-solution}
\end{equation}
where %21
\begin{equation}
D\approx 1+\frac{\kappa _{b}\varepsilon _{a0}\mu _{a}}{\kappa
_{a}\varepsilon _{b}}+2re^{-2\kappa _{b}z_{0}}.  \label{eq:Surf-dispersion}
\end{equation}
In the absence of dielectric modulations, i.e. if $\varepsilon _{a1}=0$, we
recover from (\ref{eq:Rs-solution}) the well known formula $%
R_{s}^{(0)}=-r|D|^{2}/D^{2}$, i.e. $|R_{s}^{(0)}|=1$. Obviously, $%
R_{b}^{(0)}=0$ for this case. However, from Eqs. (\ref{eq:Rs-solution}) and (%
\ref{eq:Rb-solution}), we also find the interesting result that $%
R_{s}\approx 0$ and $R_{b}\approx ir$ if the real part of the surface wave
dispersion function (\ref{eq:Surf-dispersion}) is equal to zero, i.e. %22
\begin{equation}
1+\frac{\kappa _{b}\varepsilon _{a0}\mu _{a}}{\kappa _{a}\varepsilon _{b}}+%
\frac{2(1-k_{z}^{2}\varepsilon _{b}^{2}/\kappa _{b}^{2})}{%
(1+k_{z}^{2}\varepsilon _{b}^{2}/\kappa _{b}^{2})}e^{-2\kappa
_{b}z_{0}}\approx 0,  \label{eq:Surf-DR-real}
\end{equation}
and if %23
\begin{equation}
\frac{|\varepsilon _{a1}|\omega ^{2}}{8\kappa _{a}^{2}c^{2}}=\frac{%
k_{z}\varepsilon _{b}}{\kappa _{b}(1+k_{z}^{2}\varepsilon _{b}^{2}/\kappa
_{b}^{2})}e^{-2\kappa _{b}z_{0}}.  \label{eq:Surf-DR-2-part}
\end{equation}
When $R_{s}\approx 0$ it also follows that $|R_{b}|\approx 1$. Equation (\ref
{eq:Surf-DR-real}) means that the incident wave resonantly excites a leaking
surface wave, in which most of the energy is concentrated near the interface
at $z=z_{0}$. Equation (\ref{eq:Surf-DR-2-part}) then defines the condition
for the backward energy flux to be equal to the incident flux. We note that
the relation (\ref{eq:Surf-DR-real}) only can be fulfilled if $\varepsilon
_{a0}\mu _{a}/\varepsilon _{b}<0$. For example, for a metamaterial in region 
$a$ ($\varepsilon _{a0}<0$, $\mu _{a}<0$) we must have negative dielectric
permittivity also in region $b$, whereas for a nonmagnetic material the
dielectric permittivity must change sign between the regions $a$ and $b$. In
general, to be able to study the present phenomenon, it is necessary to
consider configurations with suitable properties so that weakly damped
leaking waves can exist \cite{GradovStenflo1983}. Specifically in the
absence of the boundary region (i.e. if $z_{0}=0$, or if $\varepsilon
_{b}\rightarrow 1$), there are no such waves. In our model, it is possible
to investigate leaking waves, however \cite{GradovStenflo1983}. Similar
studies can naturally be performed for other bounded media, but the
mathematics would then be much more involved.

\paragraph{Summary}

In the present paper, we have considered the electromagnetic reflection
properties of an arbitrary dielectric and magnetic material covered by a
dielectric layer. For homogeneous media, it was shown that leaking surface
waves cannot be excited by an incident wave, and we have thus the familiar
specular reflection. For small periodic variations of the dielectric
permittivity \cite{Vukusic2006}, however, the situation is radically
different, and the coupling to leaking surface waves can then strongly
influence the reflective properties. In particular, we have found that
specular reflection is absent $(R_{s}=0)$ if the frequency of the incident
wave satisfies the dispersion relation (\ref{eq:Surf-DR-real}), and if the
width $z_{0}$ of the dielectric layer is related to the amplitude $%
\varepsilon _{a1}$ of the permittivity modulation by means of the relation (%
\ref{eq:Surf-DR-2-part}). All the incident wave energy is then reflected in
the backward direction. These properties make it possible to pick out a
specific angle and a specific frequency for a wave which is backward
reflected, while still keeping most of the signal in the ordinary reflected
signal. Obviously, such a peculiar mirror effect is of experimental interest.

%\section{References}

\end{document}